\documentclass[12pt,amstex]{article}
\usepackage{amssymb,amsmath,hyperref} 
\usepackage{latexsym}

\newtheorem{dfn}{Definition}[section] 

\newtheorem{thm}{Theorem}[section] 

\newtheorem{prop}{Proposition}[section] 
\newtheorem{lem}{Lemma}[section]

\def\cyclic{\mathop{\kern0.9ex{{+}
\kern-2.2ex\raise-.28ex\hbox{\Large\hbox
{$\circlearrowright$}}}}\limits}

\def\buildrel#1_#2^#3{\mathrel{\mathop{\kern 0pt#1}\limits_{#2}^{#3}}}

\newcommand{\Pf}{{\em Proof}. }
\newcommand{\EPf}
{%
\mbox{}%
\nolinebreak%
\hfill%
\rule{2mm}{2mm}%
\medbreak%
\par%
}

\renewcommand{\span}{\mbox{$\mathtt{span}$}}
\newcommand{\id}{\mbox{$\mathtt{Id}$}}

\newcommand{\Ad}{\mbox{$\mathtt{Ad}$}}
\newcommand{\tr}{\mbox{$\mathtt{tr}$}}
\newcommand{\ad}{\mbox{$\mathtt{ad}$}}

\newcommand{\C}{\mathbb C}
\newcommand{\E}{\mathbb E} 
\newcommand{\D}{\mathbb D}

\renewcommand{\S}{\mathbb S}

\newcommand{\R}{\mathbb R}
\newcommand{\K}{{\bf K}}

\newcommand{\g}{{\mathfrak{g}}{}} 
\renewcommand{\u}{{\mathfrak{u}}{}} 

\renewcommand{\k}{{\mathfrak{k}}{}} 
\newcommand{\p}{{\mathfrak{p}}{}} 
 
\newcommand{\n}{{\mathfrak{n}}{}} 
\newcommand{\s}{{\mathfrak{s}}{}} 
\newcommand{\ddto}{{\frac{d}{dt}|_{0}}{}}

\newcommand{\h}{{\mathfrak{h}}{}} 
\renewcommand{\a}{{\mathfrak{a}}{}} 
 
\newcommand{\z}{{\mathfrak{z}}{}}
\newcommand{\CO}{{\cal O}{}}
\newcommand{\CM}{{\cal M}{}}

\newcommand{\CS}{\mathcal S}
\newcommand{\CE}{\mathcal E}
\newcommand{\CL}{{\cal L}{}} 

\newcommand{\CC}{\mathcal C}
\newcommand{\CH}{\mathcal H}
\newcommand{\CT}{\mathcal T}
\newcommand{\CB}{\mathcal B}

\newcommand{\CF}{\mathcal F}

\newcommand{\CJ}{\mathcal J}

\newcommand{\bea}{\begin{eqnarray}}
\newcommand{\eea}{\end{eqnarray}}

\def\cref#1{Corollary~\ref{#1}}

\begin{document}

\title{Non Commutative Field Theory on Rank One Symmetric Spaces}
\author{P. Bieliavsky$^{(1)}$, R. Gurau$^{(2)}$ and V. Rivasseau$^{(2)}$\\ 
1) University of Louvain, Belgium.\\
e-mail: {\tt bieliavsky@math.ucl.ac.be}\\
2) Laboratoire de Physique Th\'eorique, CNRS UMR 8627, b\^at.\ 210\\ 
Universit\'e Paris XI,  F-91405 Orsay Cedex, France\\
e-mails: {\tt razvan.gurau@th.u-psud.fr, rivass@th.u-psud.fr}}
\maketitle 
\date{} 

\begin{abstract}
Quantum field theory has been shown recently renormalizable on flat Moyal space
and better behaved  than on ordinary space-time.
Some models at least should be completely finite, even beyond perturbation theory. 
In this paper a first step is taken to extend such theories to non-flat backgrounds
such as solvable symmetric spaces. 
\end{abstract}

\section{Introduction}

Following the pioneering work of Grosse and Wulkenhaar, noncommutative $\phi^{\star4}_4$ theory
has been shown renormalizable on four dimensional flat non-commutative Moyal space.
The first renormalization proof \cite{Grosse:2004yu} was based on 
the matrix representation of the Moyal product. It relies on adding to 
the usual propagator a marginal harmonic potential, as required by 
Langmann-Szabo duality \cite{Langmann:2002cc}, a process nicknamed "vulcanization".
An impressive explicit computation of this propagator in the matrix base
was then combined with an extensive analysis of all possible 
contractions of ribbon graphs in the RG equations  \cite{Grosse:2003aj}. 
These founding papers opened the subject of renormalizable non commutative field
theories, hereafter called RNCQFT.

The initial renormalization proof was completed and improved by introducing
multi-scale analysis, first in the matrix base \cite{Rivasseau:2005bh},
then in position space \cite{Gurau:2005gd}. The 
$\beta$-function was computed at one loop in  \cite{Grosse:2004by}, then 
shown to vanish at all orders \cite{Disertori:2006uy,Disertori:2006nq} 
at the self-duality point $\Omega=1$ (where $\Omega$ is the coefficient of the harmonic
term). The exciting conclusion is that the $\phi^{\star 4}_4$-theory is asymptotically safe,
hence free of any Landau ghost. Wave function renormalization exactly compensates
the renormalization of the four-point function, so that the flow
between the bare and the renormalized coupling is bounded.

Therefore the full non-perturbative construction of the $\phi^{\star 4}_4$-theory
should be possible. Standard constructive tools, such as cluster and Mayer expansions 
do not apply due to non-locality, but this problem can be overcome
with a new expansion called the loop-vertex expansion 
\cite{Rivasseau:2007fr}. This expansion also applies to the commutative case
\cite{Magnen:2007uy} hence provides an example
where NCQFT lead to a better understanding of \emph{ordinary} field theory.
The full construction of $\phi^{\star 4}_4$-theory now requires to extend these tools to
a multiscale analysis.

This is a vindication of the initial intuition of the founders of non commutative 
quantum field theory \cite{Schro,Heis,Snyder} that these theories should 
behave better than the ordinary ones in the ultraviolet regime.

Essentially most of the standard tools of field theory
such as parametric \cite{Gurau:2006yc,Rivasseau:2007qx}  and Mellin representations
\cite{Gurau:2007az}, dimensional regularization and renormalization \cite{Gurau:2007fy} and
the Connes-Kreimer Hopf algebra formulation of renormalization \cite{Tanasa:2007xa}
have now been generalized to RNCQFT. Other RNCQFTs have been also developed such as 
complex models, $N$-component scalar fields, and
the Fermionic NC Gross-Neveu model
and their flow computed in many cases  \cite{VignesTourneret:2006nb,VignesTourneret:2006xa,Lakhoua:2007ra,ATW,BG,BGR}. 
Their  propagator can be more complicated, namely of the \emph{covariant} type  \cite{Gurau:2005qm}, 
(studied for scalar fields in \cite{Langmann:2003cg,Langmann:2003if}),
hence describes the influence of a constant magnetic background field. Such covariant models 
are important for the future applications of RNCQFT to condensed matter problems such as 
the quantum Hall effect \cite{Suss,Poly,HellRaam}.

Concerning other scalar models, in \cite{Grosse:2005ig,Grosse:2006tc} 
the noncommutative $\phi^{\star 3}_6$-model at the  self-duality point was built and 
shown just renormalizable and exactly solvable. Self-duality  relates  
this model to the Kontsevich-model. For $\phi^{\star 3}_4$, see \cite{Grosse:2006qv}. 
The $\phi^{\star 6}_3$-model has been shown renormalizable with $x$-space 
techniques  in \cite{zhituo}. A recent review on these developments is 
\cite{Rivasseau:2007ab}.

One of the main goal for RNCQFT is to 
develop gauge theories in noncommutative geometry \cite{Connes:1994yd}. 
Yang Mills theories  are naturally obtained from the spectral action principle
\cite{Connes:1996gi,Chamseddine:1996zu} relative to an appropriate
Dirac operator. In this way, a beautiful reformulation of the standard
model of particle physics was obtained, see \cite{Chamseddine:2006ep}
for its most recent version. 

We would like RNCQFT  to connect to this reformulation, although it is also a strict extension 
of these ideas. Renormalizable theories survive RG flows, hence RNCQFTs might be the ones 
relevant to physics beyond the standard model, if non-commutativity (still confined to an "internal" space in  \cite{Chamseddine:2006ep})  fully invades space time itself at some still unexplored energy scale.

Therefore it is a central issue to find renormalizable noncommutative  gauge theories.
Usual Yang-Mills theory on Moyal space is known to be not
renormalisable \cite{Matusis:2000jf}. 
The principal of spectral action is to compute a one-loop effective action of Fermions in a
classical external gauge field. Translated from Fermions to scalar fields
as in \cite{Gayral:2004cs}, the computation was completed
in position space \cite{deGoursac:2007gq} 
and in the matrix base \cite{Grosse:2007dm}. 
The problem with such effective actions is that $A_\mu=0$ \emph{is not a stable solution of the classical field equation}.  Recently a particular non-trivial explicit vacuum was found under a radial Ansatz, as part of a more general study of
a Connes-Lott model  \cite{Connes:1990qp} (Yang-Mills theory coupled to the Higgs field) but in which 
the spectral action is derived from a non-standard Dirac operator, namely the 
square root of the Grosse-Wulkenhaar propagator. In this way the relation 
between the Higgs mechanism, the non-trivial gauge 
vacuum and the harmonic potential for the Higgs field is somewhat clarified
\cite{Grosse:2007jy}. 


Parallel to these developments the mathematical theory of Moyal products was generalized to a category
of spaces called the solvable symmetric spaces \cite{Bi07}. At the origin was a question 
raised by A. Weinstein in the context of deformation quantization of symmetric spaces. Roughly speaking, the so-called `WKB quantization program' for symmetric spaces may be described as follows.

An affine manifold $(M,\nabla)$ is called {\sl symmetric} if 
the local geodesic symmetry centered at any point $x$ of $M$
globally extends as an affine transformation of $(M,\nabla)$:
\begin{eqnarray*}
s_x:M&\longrightarrow& M\\
\mbox{\rm Exp}_x^{-1}\,\circ \,s_x\,\circ\,\mbox{\rm Exp}_x&=&-\mbox{id}\;.
\end{eqnarray*}
It is called a {\sl symplectic symmetric space} or SSS if the symmetries preserve a non-degenerate 
two-form $\omega$ (which in this case is then automatically closed). On such a space which is strictly 
geodesically convex (e.g. a Hermitian symmetric space of the non-compact type), one 
defines the following three-point function:
\begin{equation}\label{DT}
\Phi^{-1}: M\times M\times M\longrightarrow M\times M\times M: (x_1,x_2,x_3)\,\mapsto\,(\,m(x_1,x_2)\,,\,m(x_2,x_3)\,,\,m(x_3,x_1)\,)\;,
\end{equation}
where $m(x,y)$ denotes the mid-point on the geodesic line between $x$ and $y$ $\ ( s_{m(x,y)}x=y)$.
The latter function admits an inverse $\Phi$ locally around the diagonal in $M^3$. Therefore, when there is no
second de Rham cohomology and for `small geodesic triangles', one may define the following local function:
\begin{equation*}
S_W\;:=\; \mbox{\rm Area}\,\circ\,\Phi\;,
\end{equation*}
where, given an oriented geodesic triangle $\stackrel{\Delta}{xyz}$ in $M$, one sets:
\begin{equation*}
 \mbox{\rm Area}(x,y,z)\;:=\;\int_{\stackrel{\Delta}{xyz}}\,\omega\;.
\end{equation*}
Now, given a symmetry invariant formal star product $\star$ on $(M,\omega)$ that admits the following `WKB' form:
\begin{equation*}
u\star v(x)\;=\;\frac{1}{\theta^{2n}}\,\int_{M\times M} A_\theta(x,y,z)\,e^{\frac{i}{\theta} S(x,y,z)}\,u(y)\,v(z)\,\omega^n_y\otimes\omega^n_z\;,
\end{equation*}
for some invariant (formal) amplitude  $A_\theta$ and phase $S$, A. Weinstein  suggests in \cite{Wein} that
the above function $S_W$ should be relevant in the asymptotics of the phase :
\begin{equation*}
S\;\sim\;S_W\;.
\end{equation*}

The problem was thus to develop a suitable geometrical framework for the study of asymptotic
quantization of symplectic  symmetric spaces allowing to decide about Weinstein's asymptotics
as well as to describe the amplitude $A_\theta$ within a non-formal context.

Two main classes have been treated in the solvable situation (i.e. when the transvection group of the 
symmetric space is a solvable Lie group): the class of {\sl elementary spaces} \cite{Bi07a} and the class
of {\sl rank one spaces} \cite{BiMs01}. In those cases, the WKB-quantization program has been carried out explicitly and within the non-formal setting. In particular, the phase function $S$, in the solvable case, appears to coincide with 
Weinstein function $S_W$ up to a possible extra boundary term. 

It is now a natural question to ask whether RNCQFT and vulcanization extend to  non-trivial geometric backgrounds,
as RNCQFT might eventually be applied in the presence of strong gravitational fields e.g. in  the vicinity 
of a black hole horizon. In order to have at least semi-infinite renormalization flows 
we need an infinite-dimensional Moyal algebra of functions. This rules out compact 
spaces such as fuzzy spheres but points to the interesting case of SSS.

This paper is organized as follows:

Section 2 recalls the definitions and 
basic properties of the Moyal product on SSS.
In section 3 we extend the notion of Langmann-Szabo duality to SSS
and compute the vulcanized Laplacian explicitly.

In section 4 we give the action functional of the $\phi^{\star 4}_{4}$ theory on SSS.
and  we show that the properties of associativity and traceability
of the Moyal products on SSS allow one to extend to these backgrounds the theory 
of Filk moves \cite{Filk1996dm} and the "Moyality" of Feynman amplitudes for planar graphs
 with a single external face \cite{Rivasseau:2007ab}.

This completes essentially the model-independent steps
of the renormalization program, as it generalizes to SSS 
the dimension-independent locality principle of ordinary geometry.

In section 5 as a prelude to the spectral analysis of the propagator which is an essential
step for renormalization, we remark that the vulcanized Laplacian kernel
is a generator of a representation of the metaplectic group. This should allow
access to its spectral properties, and to explicit bounds on this kernel.

In conclusion we comment briefly on the remaining steps for a full renormalization
proof of vulcanized $\phi^{\star 4}_4$ on the simplest SSS space of dimension 4, 
e.g. $SU(1,2)/U(2)$, which is devoted to future publications.

\medskip
\noindent{\bf Acknowledgments}
\medskip

P.B.  acknowledges partial support from the IAP grant `NOSY' delivered by the Belgian Federal Government 
as well as partial support from the IISN-Belgium (convention 4.4511.06).
R. G and V. R. acknowledge the support of the ANR-Genophy grant.

\section{Definitions and notations}
\subsection{Hermitian symmetric spaces of the non-compact type}\label{HSS}
The so-called `Moyal plane' corresponds to the deformation quantization ($\star$-product) 
of the  (flat) Euclidean vector space $(V:=\R^{2n},\beta^0)$ endowed with its canonical bilinear symplectic
2-form $\omega^0=\sum_{j=1}^n{\rm d}p_j\wedge{\rm d}q_j$. In order to pass to a curved framework,
one is, on the first hand, naturally led to consider the class of Riemannian manifolds $(M,\beta)$ endowed
with a {\sl compatible} symplectic structure $\omega$. By {\sl compatible}, we mean that the triple $(M,\beta,\omega)$ is a K\"ahler manifold. On the second hand, the use of geodesic symmetries within quantization processes has been shown to be very efficient \cite{Berezin, Unterberger, Wein, Bi07a}. At last, Sekigawa and Vanhecke showed in \cite{Seki} that a K\"ahler manifold whose symplectic structure is invariant by the geodesic symmetries must be a Hermitian symmetric space i.e. the 
geodesic symmetries preserve the Levi-Civita connection as well.

 Another important feature of the Moyal plane is its {\sl traciality}:
the Hilbert space of square integrable functions $\CH:=L^2(\R^{2n})$ is stable
under Weyl's product $\star^W$ and turns out to be a Hilbert algebra in the sense that
the following property holds:
\begin{equation}\label{TRAC}
\int a\star^Wb\;:=\;\int ab\qquad(a,b\in\CH).
\end{equation}
In the curved compact situation, however, there is no hope that such a property would hold at a non-formal level
\cite{Rieffel} as shown by analyzing the simplest case of the 2-sphere.

 These considerations therefore lead us to focus on the class of Hermitian symmetric spaces of the
non-compact type, which we present here only in the rank one case (the higher rank situation can be treated by
applying the `extension lemma' in \cite{BCSV}).
\begin{dfn}
A rank one Hermitian symmetric space of the non-compact type is a homogeneous manifold $SU(1,n)/U(n)$ ($n\geq1$).
\end{dfn}

The metric and symplectic structures may be described as follows. Denoting $G:=SU(1,n)$ and $K:=U(n)$,
the Lie algebra $\g$ of $G$ splits into a direct sum of vector spaces $\g=\k\oplus\p$ where $\k$
is the Lie algebra of the maximal compact subgroup $K$ and where $\p$ is its orthonormal complement
with respect to the Killing form $B$ on $\g$. The above decomposition may also be seen as the ($\pm1$)-eigenspace decomposition with respect to the Cartan involution $\sigma$ of $\g$ associated with the choice
of $K$. Note that in this case the Cartan involution is interior: there exists an element ${\bf j}$ in the center
$Z(K)=U(1)$ of $K$ such that $\sigma=\Ad({\bf j})$. The differential at the unit element $e$ of the
projection $\pi:G\to G/K$ identifies the vector space $\p$ with the tangent space $T_K(\D)$ at $K$ to the manifold $\D:=G/K$. The vector space $\p$ is an invariant subspace under the (adjoint) action of $K$ on $\g$. Moreover, the tensors $B|_{\p\times\p}$, ${\cal J}:=\ad(\log{\bf j})|_\p$ and $\omega:=B|_{\p\times\p}\circ\id\otimes{\cal J}$ are 
invariant under the $K$-action. They globalize to $\D$ as respectively the metric $\beta$, the complex structure
$J$ and the symplectic (K\"ahler) 2-form $\omega$. The geodesic symmetry $s^\D_{gK}$ centered at $gK$ with respect
to the Levi-Civita connection $\nabla^\D$ of $\beta$ is explicitly described by the following expression:
\begin{equation*}
s^\D_{gK}(hK)\;=\;g{\bf j}g^{-1}h{\bf j}^{-1}K\;.
\end{equation*}
A second picture of the geometry of $\D$ relies on the (co) adjoint representation. Indeed,
the adjoint orbit $\CO$ of $Z:=\log({\bf j})$ turns out to be isomorphic as a $G$-homogeneous space
to $\D$. The symplectic structure identifies with the KKS-form:
\begin{equation*}
\omega_x(X^\star,Y^\star)=B(x,[X,Y])\;,
\end{equation*}
where $x\in\CO\subset\g$ and where $X^\star$ denotes the fundamental vector field $X^\star_x:=\frac{d}{dt}|_0\Ad(\exp(-t X))x$ associated to $X\in\g$. The metric is then simply the induced metric on $\CO$
from the Killing form $B$ on $\g$.

 A third and last picture is based on the Iwasawa decomposition of $G$. Recall that all Abelian subalgebras of $\g$
contained in $\p$ are $K$-conjugated. Let $\a\subset\p$ be such an Abelian subalgebra. Its adjoint action
is then semisimple and on gets a decomposition of $\g$ into root spaces:
\begin{equation*}
\g=\g_0\oplus\bigoplus_{\alpha\in\Phi}\g_\alpha
\end{equation*}
where for $\alpha\in\a^\star$, one sets $\g_\alpha:=\{X\in\g|[H,X]=\alpha(H)X\quad(H\in\a)\}$ and where
$\Phi$ is the subset of $\a^\star\backslash0$ for which $\g_\alpha$ is not reduced to zero.
A choice of an orthonormal basis of $\a$ yields a partition of the set $\Phi$ into $\Phi^+\cup\Phi^-$
with $\Phi^-=-\Phi^+$ such that the subspace
\begin{equation*}
\n:=\bigoplus_{\alpha\in\Phi^+}\g_\alpha
\end{equation*}
is a (nilpotent) subalgebra of $\g$. In the rank one case with $n\geq2$, one has $\dim\a=1$, $\Phi^+=\{\alpha,2\alpha\}$, 
$\dim\g_{2\alpha}=1$ and $\n$ is isomorphic to a Heisenberg algebra\footnote{Note that $\alpha$
can be normalized to 1.}. The direct sum $\s=\a\oplus\n$
is then a solvable subalgebra of $\g$ whose associated simply connected Lie group $\S$
is an exponential $H$-group (one dimensional split extension of a Heisenberg group).
The map
\begin{equation}\label{DARBOUX}
\s=\a\oplus\g_\alpha\oplus\g_{2\alpha}\longrightarrow\D:(a,x,z)\mapsto\exp(a)\exp(x)\exp(z)K
\end{equation}
turns out to be a global Darboux diffeomorphism when $\s$ is endowed with the bilinear
form $\omega = {\rm d}a\wedge{\rm d}z+\omega^0$ where $\omega^0$ denotes the symplectic form
on $\g_{\alpha}$ inducing the bracket in the Heisenberg algebra $\n$. Note that the above map
define global coordinates on the group $\S$ on which the 2-form $\omega$ is left-invariant. The identifications $\S=\CO=\D$ are then 
$\S$-equivariant symplectic identifications.
\subsection{Contracted spaces}
The symplectic Lie group $\S$ is `of symmetric type' in the sense that it carries a symmetric affine connection
($\nabla^\D$) which is left-invariant. However, there exists on $\S$ another symmetric affine connection, $\nabla^c$, left-invariant and moreover symplectic (i.e. $\nabla^c\omega=0$). This connection is curved and is not
the Levi-Civita connection associated to any Riemannian or pseudo-Riemannian metric. It is a purely 
symplectic connection. Below we describe some features of its associated geometry. 
One defines the smooth map  $s:\S\times\S\to\S:(x,y)\mapsto s_x(y)$ given in coordinates $(a,x,y)$, by
\begin{equation*}
s_{(a,x,z)}(a',x',z')=(2a-a',2  \cosh(a-a')x-x',
   			2  \cosh(2 (a -  a'))z + \omega^0(x,x') \sinh(a - a')-z').
\end{equation*}

One then observes that it is $\S$-equivariant (i.e. $xs_yx^{-1}z=s_{xy}z$ for all $x,y,z\in\S$).
Moreover for every $x$, the map $s_x$ is an involutive symplectomorphism of $(\S,\omega)$.
At last, one has the identity: $s_xs_ys_x=s_{s_x(y)}$. The following formula defines an affine connection on $\S$:
\begin{equation}\label{CONNECTION}
\omega_x(\nabla^c_XY,Z)\;:=\;\frac{1}{2}\,X_x.\omega(Y+s_{x_\star}Y\,,\,Z)\;,
\end{equation}
where $X,Y,Z$ are any vector fields on $\S$. The connection $\nabla^c$ turns out to be torsion-free, symplectic,
and is the only affine connection which is invariant under the involutions $\{s_x\}_{x\in\S}$.
The latter involutions actually coincide with the geodesic symmetries associated to $\nabla^c$ i.e. the
triple $(\S,\omega,s)$ is a
{\sl symplectic symmetric space} (\cite{Bi07, BCG95}). We call it the {\bf  contracted space}. Its {\sl transvection group}
$G^c$ generated by even products of symmetries is a transitive solvable Lie group of affine symplectomorphisms
of $(\S,\omega,s)$. It plays the role of $G$ in the non-contracted situation of $\D$. Similarly as $G$, the group
$G^c$ contains $\S$ as a subgroup.
\subsection{WKB-quantization}
\begin{thm}\label{BIMAS}
For all non-zero $\theta\in\R$, there exists a Fr\'echet function space 
$\CE_\theta$,                                                           
$C^\infty_c(\S)\subset\CE_\theta\subset C^\infty(\S)$,   
such that, defining for all $u,v\in C^\infty_c(\S)$                  
\begin{eqnarray}  \label{PRODUCT}
\nonumber
&&(u\star_\theta v)(a_0,x_0,z_0) \\  \nonumber
 &&= \frac{1}{\theta^{\dim \S}}
\int_{ \S\times \S}
\cosh(2(a_1-a_2))\,[\cosh(a_2- a_0)\cosh(a_0-a_1)\,]^{\dim\S-2}\\  \nonumber
 && \times\exp\Big( \frac{2i}{\theta}\Big\{ S^0\big(\cosh(a_1-a_2)x_0, 
\cosh(a_2-a_0)x_1, \cosh(a_0-a_1)x_2\big)\\ 
 &&- \cyclic_{0,1,2}\sinh(2(a_0-a_1))z_2 \Big\} \Big) u(a_1,x_1,z_1)\,v(a_2,x_2,z_2)\, da_1da_2dx_1dx_2dz_1dz_2 
\end{eqnarray}
where $S^0(x_0,x_1,x_2):=\omega^0(x_0,x_1)+\omega^0(x_1,x_2)+\omega^0(x_2,x_0)$ 
is the phase for the Weyl product on $C^\infty_c(\g_\alpha)$ and                            
$\cyclic_{0,1,2}$ stands for cyclic summation, one has: 

\begin{enumerate}
\item  $u\star_\theta v$ is smooth and the map                                    
$ C^\infty_c(\S)\times C^\infty_c(\S)
\to C^\infty(\S)$
extends to an associative product on $\CE_\theta$.                                          
The pair $(\CE_\theta,\star_\theta)$ is a (pre-$C^\star$) Fr\'echet algebra.
\item In coordinates $(a,x,z)$ the group multiplication law reads
\[ 
L_{(a,x,z)}(a',x',z')=\left(
a+a',e^{-a'}x+x',e^{-2a'}z+z'+\frac{1}{2}\omega^0(x,x')e^{-a'}
\right).
\]
The phase and amplitude occurring in formula (\ref{PRODUCT}) are both invariant 
under the left action $L:\S\times \S\to \S$.
\item Formula (\ref{PRODUCT}) admits a formal asymptotic expansion of the form:
\begin{equation*}
u\star_\theta v\sim \,uv\,+\,\frac{\theta}{2i}\{u,v\}\,+O(\theta^2)\;;    
\end{equation*}
where $\{\,,\,\}$ denotes the symplectic Poisson bracket on 
$C^\infty(\S)$ associated with $\omega$.
The full series yields an associative formal star product on 
$(\S,\omega)$ denoted by $\tilde{\star}_\theta$.
\end{enumerate}
\end{thm}
Roughly speaking, the above product can be obtained by intertwining the standard Weyl-Moyal product
$\star^0_\theta$ on the Schwartz space $\CS(\s)=:\CS$ (endowed with the Poisson structure associated with
the two-form $\omega$). Indeed, one formally has:
\begin{equation*}
u\star_\theta v\;=\; T_{\theta,1}\,(\,T^{-1}_{\theta, 1}u\,\star^0_\theta\,T^{-1}_{\theta,1}v\,)\;;
\end{equation*}
where, denoting by 
\begin{equation}\label{partialFourier}
\CF_\z u(a,v,\xi)\;:=:\;\hat{u}(a,v,\xi):=\int e^{-i\xi z}u(a,v,\ell)\,{\rm d}\ell
\end{equation}
the partial Fourier transform in the $z$-variable, and defining  the following smooth one-parameter family of diffeomorphisms:
\begin{equation*}
\phi_\theta:(a,v,\xi)\mapsto(a,\frac{1}{\cosh(\theta\xi)}v,\frac{1}{2\theta}\sinh(2\theta\xi))\;;
\end{equation*}
the $T_{\theta,1}$ map is defined as 
\begin{equation}\label{TAUS}
T_{\theta,1}\;:=\;\CF_\z^{-1}\circ(\phi^{-1}_\theta)^\star\circ\CF_\z\,:\,\CS\to\CS'\;.
\end{equation}
The space $\CE_\theta$ then corresponds to the range of the above map in the tempered distributions $\CS'$.

In dimension two ($n=1$), the space $SU(1,1)/U(1)$ is isometric to the hyperbolic plane. The algebra
$\n$ is then one dimensional ($\g_\alpha$ disappears) and the group $\S$ is isomorphic to the affine group $ax+b$ whose group law, in coordinates $(a,z)$, reads $(a,z).(a',z')=(a+a',e^{-2a'}z+z')$. The 
product formula (\ref{PRODUCT}) degenerates as 
\begin{eqnarray*}
(u\star_\theta v)(a_0,z_0)&=&\\
\frac{1}{\theta^{2}}
\int_{ \S\times \S}
\cosh(2(a_1-a_2))&&\hspace{-7mm}\exp\Big( \frac{-2i}{\theta}
\cyclic_{0,1,2}\sinh(2(a_0-a_1))z_2 \Big)\;
u(a_1,z_1)\,v(a_2,z_2)\, da_1da_2dz_1dz_2\;.
\end{eqnarray*}
Again it can be seen as transporting Weyl-Moyal's product by a intertwiner of the above form (\ref{TAUS}) with 
$\phi_\theta(a,\xi):=(a,\frac{1}{2\theta}\sinh(2\theta\xi))$.

\subsection{Traciality}
The above products are not tracial in the sense that $L^2(\S)$ is stable and that property (\ref{TRAC}) holds. 
Nevertheless, they can be easily modified into  strongly tracial products \cite{Bi07c}.
\begin{prop}
Let $\theta>0$ and denote by $L^2(\S)$ the space of functions on $\S$ that are square integrable 
w.r.t. a left-invariant Haar measure. Set
\begin{equation*}
S\;:=\;S^0\big(\cosh(a_1-a_2)x_0, 
\cosh(a_2-a_0)x_1, \cosh(a_0-a_1)x_2\big) - \cyclic_{0,1,2}\sinh(2(a_0-a_1))z_2\;;
\end{equation*}
for the phase function as in the product formula (\ref{PRODUCT}).
Let $u,v\in C^\infty_c(\S)$. Then the following formula:
\begin{equation}\label{starcan}
u\star^{\mbox{\tiny{\rm can}}}_\theta v\;=\;\frac{1}{\theta^{\dim \S}}\int_{\S\times \S}\,{\bf A}_{\mbox{\rm can}}\;e^{\frac{2i}{\theta}S}\;u\otimes v\;;
\end{equation}
where
\begin{eqnarray*}
{\bf A}_{\mbox{\rm can}}(x_0,x_1,x_2)&:=&
\sqrt{\cosh(2(a_0-a_1))\cosh(2(a_1-a_2))\cosh(2(a_2-a_0))}\,\times \\ &\times&[\cosh(a_0-a_1)\cosh(a_1- a_2)\cosh(a_2- a_0)\,]^{\frac{\dim\S-2}{2}}\;,
\end{eqnarray*}
extends to a complex bilinear associative product $\star^{\mbox{\tiny{\rm can}}}_\theta:L^2(\S)\times
L^2(\S)\to L^2(\S)$. The pair $(L^2(\S),\star^{\mbox{\tiny{\rm can}}}_\theta)$ is a Hilbert algebra
on which the symmetries $\{s_x\}_{x\in\S}$ act by isometries. Moreover, the traciality property 
(\ref{TRAC}) holds.
\end{prop}
 In dimension two, one has (\cite{StarP,Bi07a})
\begin{prop}
Let $\theta>0$ and denote by $L^2(\S)$ the space of functions on $S$ that are square integrable 
w.r.t. a left-invariant Haar measure. Let $u,v\in C^\infty_c(\S)\qquad(n=1)$. Then the following formula:
\begin{eqnarray}
&&(u\star^{\mbox{\tiny{\rm can}}}_\theta v)(a_0,z_0)=\nonumber 
\frac{1}{\theta^{2}}
\int_{ \S\times \S}
\left[\cosh(2(a_0-a_1))\cosh(2(a_1-a_2))\cosh(2(a_2-a_0))\right]^{\frac{1}{2}}\\
&&\hspace{2cm} \times\exp\Big( \frac{-2i}{\theta}
\cyclic_{0,1,2}\sinh(2(a_0-a_1))z_2 \Big)\;
u(a_1,z_1)\,v(a_2,z_2)\, da_1da_2dz_1dz_2\;,
\end{eqnarray}
extends to a complex bilinear associative product $\star^{\mbox{\tiny{\rm can}}}_\theta:L^2(\S)\times
L^2(\S)\to L^2(\S)$. The pair $(L^2(\S),\star^{\mbox{\tiny{\rm can}}}_\theta)$ is a Hilbert algebra
on which the symmetries $\{s_x\}_{x\in\S}$ act by isometries. Moreover, the traciality property 
(\ref{TRAC}) holds.
\end{prop}
These tracial products may be seen as transporting Weyl-Moyal by a modified intertwiner of the form:
\begin{equation*}
T_{\theta,{\mbox{\tiny{\rm can}}}}\;:=\;\CF_\z^{-1}\circ\CM_{\sqrt{\mbox{\rm Jac}_{\phi_\theta^{-1}}}}\circ(\phi^{-1})^\star_\theta\circ\CF_\z\;;
\end{equation*}
where $\CM_f$ denotes the multiplication operator by $f$. The above map 
actually defines a unitary automorphism:
\begin{equation*}
T_{\theta,{\mbox{\tiny{\rm can}}}}\,:\, L^2(\S)\longrightarrow L^2(\S)\;,
\end{equation*}
where we identified $L^2(\S)$ with $L^2(\s)$ via the global Darboux chart (\ref{DARBOUX}).

\section{LS-duality}

\subsection{Definition}

Within the flat context, Langmann-Szabo duality coincides with the symplectic Fourier transform. It plays a key role
in making the corresponding field theory renormalizable (perturbatively and probably also
at the constructive level, see the introduction). 

On curved Hermitian symmetric spaces we have natural notions of Laplacian and 
of invariant star product but LS duality has to be defined. In this section we make a 
first proposal in this direction: simply transport the ordinary LS duality in the flat case 
to the curved case under the $T$ transformation. Then we consider the curved Laplacian
(which is certainly not spectrally equivalent to the flat one) and extend it into an LS
covariant operator by adding $\Omega^2$ times  its LS dual.
This procedure mimics the Grosse-Wulkenhaar one but does
not lead to an equivalent theory simply because when the Grosse-Wulkenhaar $\Omega$
parameter is turned off, the limits are different.

More precisely, on the  space $\CS$ of Schwartz functions defined on the symplectic vector space $(\s,\omega:={\rm d}a\wedge{\rm d}z+\omega^0)$, one defines the following integral transformations:
\begin{equation*}
\CF_{\pm\omega}(u)(x)\;:=\;\int_\s\,e^{\mp i\omega(x,y)}\;u(y)\,{\rm d}y\;.
\end{equation*}
These then define topological automorphisms of $\CS$ extending as unitary isomorphisms to $L^2(\s)$.
\begin{dfn}
The {\bf ($\pm$) LS-transform} on $L^2(\S)$ is the involutive unitary automorphism defined as 
\begin{equation*}
\CF_{\mbox{\tiny{\rm can}}}^{\pm}\;:=\;T_{\theta,{\mbox{\tiny{\rm can}}}}\,\circ\,\CF_{\pm\omega}\,\circ\,T_{\theta,{\mbox{\tiny{\rm can}}}}^{-1}\;.
\end{equation*}
\end{dfn}
Our  intertwiner may be expressed in terms of the flat LS-transform, indeed the following lemma holds
by a straightforward computation.
\begin{lem} Denote by $\Phi_\theta:\S\to\S$ the following one parameter family of diffeomorphisms:
\begin{equation*}
\Phi_\theta(a,v,z)\;:=\;(\,\frac{1}{2\theta}\sinh(2\theta a)\,,\,\cosh(\theta a)\,v\,,\,z\,)\;.
\end{equation*}
Then, one has
\begin{equation*}
T_{\theta,1}^{-1}\;=\;\CF_{\omega}\circ\CM_{[\cosh(\theta a)]^{\dim\S-2}}\circ\Phi^\star_\theta\circ\CF_{\omega}\;.
\end{equation*}
\end{lem}

\subsection{The Laplacian}
We first describe the standard Laplace operator on $G/K$ identified with $\S$. As we have seen, the Laplace operator $\Delta$ is a second order left-invariant differential operator on the Lie group $\S$. Moreover its principal symbol coincides with the quadratic form induced by the invariant metric $\beta$ on $G/K$. It is therefore of the form $B_{ij}\tilde{X}_i\tilde{X}_j+B_i\tilde{X}_i+B$,
where $\{X_i\}$ is a basis of $\s$ and where $B_{ij}=B({X_i}_\p,{X_j}_\p)$.  Note that for all $X,Y\in\g$, since $Z$ is central in $\k$ and orthogonal to $\p$, the invariance of the Killing form yields: $\omega(X_\p,Y_\p)=B(Z,[X,Y])$. In the same way, one gets $B(X_\p,Y_\p)=B(Z,[[Z,X],Y])$.
Remark also that $X_\p=-[Z,[Z,X]]$. Now, $B(H,\n_\p)=B(Z,[[Z,H],\n])=B([Z,[Z,H]],\n)=B(H,\n)=0$.
Similarly, $B({\g_{2\alpha}}_\p,{\g_\alpha}_\p)=B(Z,[[Z,\g_{2\alpha}],{\g_\alpha}_\p])=B({\g_{2\alpha}}_\p)=B(\g_{2\alpha}-\sigma\g_{2\alpha},\g_{\alpha})=B(\g_{2\alpha}+\g_{-2\alpha},\g_{\alpha})=0$. In particular,
since $\a\oplus\g_{2\alpha}$ and $\g_{\alpha}$ are symplectic, there are also stable
under the complex structure $\CJ$. Fix then a Lagrangian subspace $\CL$ of $\g_{2\alpha}$ in
duality with $\CL':=\CJ(\CL)$. Consider an orthonormal basis $\{e_k\}$ of $\CL$ as well as its 
dual basis $\{f_k:=\CJ(e_k)\}$ in $\CL'$.
Observe that the left-invariant vector fields are expressed as:
\begin{eqnarray*}
\tilde{H}&=&\ddto\exp(aH)\exp(x)\exp(zE)\exp(tH)=\ddto\exp((a+t)H)\exp(e^{-t}x)\exp(ze^{-2t}E)=\\
&=&\partial_a-x^j\partial_{x^j}-2z\partial_z\,;\\
\tilde{v}&=& \ddto\exp(aH)\exp(x+tv)\exp((z+\frac{1}{2}t\omega(x,v))E)=\partial_v+\frac{1}{2}\omega(x,v)\partial_z\,;\\
\tilde{E}&=&\partial_z\,.
\end{eqnarray*}
In particular, the fields $\tilde{v}$ as well as $\tilde{E}$ preserve the Liouville volume.
By imposing self-adjointness, one observes that the Laplace operator admits the expression:
\begin{equation*}
\Delta=\tilde{H}^2+\sum_k(\tilde{e_k}^2+\tilde{f_k}^2)+\tilde{E}^2-(\dim\S)\tilde{H}\;;
\end{equation*}
where we adopt the normalization:
$\omega(H,E)=2B(Z,E):=1$.

 For instance, in the $n=2$ case, one has the following matrix representation:
\begin{eqnarray*}
\sigma A=-A^\dagger\ , \quad  
\k =\;\{ &  \left(  \begin{array}{cc}-\tr X&0\\ 0&X\end{array}\right)\;|\;X\in\u(2)\} \ ,&
\p\;=\;\{\left(\begin{array}{cc} 0&v^\dagger\\ v&0\end{array}\right)\;|\;v\in\C^2\}\\
Z\;=\;\frac{1}{3}\left(\begin{array}{cc}-2i&0\\ 0&iI\end{array}\right)\ ,&
H \;=\;\left(\begin{array}{ccc}0&1&0\\ 1&0&0\\ 0&0&0\end{array}\right)\ ,&
E \;=\;\left(\begin{array}{ccc}i&-i&0\\ i&-i&0\\ 0&0&0\end{array}\right)\ ,\\
\xi\;=\;\left(\begin{array}{ccc}0&0&\overline{\xi}\\ 0&0&\overline{\xi}\\ \xi&-\xi&0\end{array}\right)\ ,&
E_\p\;=\;\left(\begin{array}{ccc}0&-i&0\\ i&0&0\\ 0&0&0\end{array}\right)\ ,&
\xi_\p\;=\;\left(\begin{array}{ccc}0&0&\overline{\xi}\\ 0&0&0\\ \xi&0&0\end{array}\right)\ ,
\end{eqnarray*}
where $\xi:=:xe_1+yf_1=:x+iy$ and ${}^\dagger$ stands for transposed conjugated.

In order to define a propagator that is invariant under the LS-transform, we first recall some facts. 
View the space  $\S$ as the adjoint orbit $\CO$ in $\g$ endowed with the KKS form $\omega$
as described in Section \ref{HSS}. Consider $(\CO,\omega,G)$ as a strongly Hamiltonian system and 
denote by
\begin{equation*}
\lambda:\g\to C^\infty(\CO):X\mapsto\lambda_X
\end{equation*}
the associated classical moment mapping (i.e. $\lambda_X(x):=B(x, X)$). Express the latter within
Iwasawa coordinates (\ref{DARBOUX}) as 
$\lambda_X(a,v,z):=\lambda_X(\exp(aH)\exp(v+zE) Z)$.
On then has 
\begin{prop}\cite{BiMas02}
The Moyal formal star product $\star^0_\nu$ on $(\s, \omega:={\rm d}a\wedge{\rm d}z+\omega^0)$
is $\s$-covariant in the sense that for all $X,Y\in\s$, one has:
\begin{equation*}
[\lambda_X,\lambda_Y]_{\star^0_\nu}\;=\;2\nu\lambda_{[X,Y]}\;.
\end{equation*}
\end{prop}
At the formal level, this yields a representation of $\s$ acting on $\E_\nu:=C^\infty(\S)[[\nu]]$ by derivations of the
Moyal product:
\begin{eqnarray*}
\rho_\nu\;:\;\s\longrightarrow\;\mathfrak{Der}(\E_\nu)\,:\\
\rho_\nu(X)u\;:=\;\frac{1}{2\nu}\,[\,\lambda_X\,,\,u\,]_{\star^0_\nu}\;.
\end{eqnarray*}
From \cite{BiMas02}, one finds the following expressions:
\begin{eqnarray*}
\rho_\nu(E)u&=&\frac{-1}{\nu}e^{-2a}\sinh(2\nu\partial_z)\,u\\
\rho_\nu(w)u&=&e^{-a}\left(\cosh(\nu\partial_z)\partial_wu-\frac{1}{\nu}\omega(v,w)\sinh(\nu\partial_z)u\right)\\
\rho_\nu(H)u&=&-\partial_au\;.
\end{eqnarray*}
One then has
\begin{prop}\label{XSTAR} Let $X\in\s$ and denote by $X^\star$ the associated right-invariant
vector field on the group $\S$. Then:
\begin{equation*}
T_{\theta,{\mbox{\tiny{\rm can}}}}\,\circ\,\rho_\nu(X)\,\circ\,T_{\theta,{\mbox{\tiny{\rm can}}}}^{-1}\;=\;X^\star\;.
\end{equation*}
\end{prop}
\Pf
We have seen that $T_{\theta,{\mbox{\tiny{\rm can}}}}=\CC\circ T_{\theta,1}\circ\CC^{-1}$ where  $\CC:=\CF^{-1}_\z\CM_P\CF_\z$ for some complex-valued one variable function $P$. Proposition 2.2 of \cite{BiMas02} tells us that $T_{\theta,{\mbox{\tiny{\rm can}}}}\,\circ\,\rho_\nu(X)\,\circ\,T_{\theta,{\mbox{\tiny{\rm can}}}}^{-1}=\CC\circ X^\star\circ\CC^{-1}$. Observing that the operator $\CC$
is nothing else than a left-invariant convolution operator on the Lie group $\S$ \cite{Bi07new} and since $X^\star$ is the infinitesimal generator of the left multiplication on $\S$, one obtains the result. \EPf

{\underline{Chirality}}: The scalar Laplacian operator on $\D=\S$  is a left-invariant differential operator.
It therefore may be expressed within a basis $\{X_j\}$ of $\s$ as $\Delta^L=\beta_{jk}\tilde{X_j}\tilde{X_k}+\beta_j\tilde{X_j}+\beta$ where $\beta_{jk},\beta_j$ and $\beta$ are real constants
and where $\tilde{X}$ denotes the left-invariant vector field on $\S$ associated to $X\in\s$.

 In the same way, one has a right-invariant operator $\Delta^R$ expressed in terms of the right-invariant vector fields $X_j^\star$ associated with a right-invariant Riemannian metric on $\S$. Of course, intrinsically,
the left and right-geometries on $\S$ are isometric, essentially under the inversion map.

 Conventionally, we choose to consider the {\bf right-invariant Laplacian} together with the {\bf left-invariant star
products} on $\S$.  Proposition \ref{XSTAR} then yields the following expression for $\Delta^R$:
\begin{equation*}
\Delta^R\,u\;=\;\frac{1}{4\nu^2}\beta_{jk}[[\,\Lambda_{X_j}[\,\,\Lambda_{X_k}\,,\,u]_{\star_\nu}\,]_{\star_\nu}\,
\,+\,\frac{1}{2\nu}\beta_j\,[\,\Lambda_{X_j}\,,\,u\,]_{\star_\nu} \,+\,\beta\,u\;;
\end{equation*}
where $\star_\nu$ denotes the $\S$-invariant {\sl formal} star product intertwined with $\star^0_\nu$ under (the formal asymptotic expansion of) $T_{\theta,{\mbox{\tiny{\rm can}}}}$; and, where $\Lambda$ denotes the associated intertwined quantum moment (see \cite{BBM} for details about the latter asymptotic expansion).

\subsection{LS dual of the Laplacian}\label{LSDUAL}

 In view of LS-invariance, one is therefore left with considering the following formal operator for every $X$ in $\s$:
\begin{equation*}
u\;\mapsto \;\CF_{\mbox{\tiny{\rm can}}}^{\pm}\,\circ\,T_{\theta,{\mbox{\tiny{\rm can}}}}\,\circ\,\rho_\nu(X)\,\circ\,T_{\theta,{\mbox{\tiny{\rm can}}}}^{-1}\,\circ\,\CF_{\mbox{\tiny{\rm can}}}^{\pm}\,(u)\;.
\end{equation*}
Equivalently:
\begin{equation*}
\CF_{\mbox{\tiny{\rm can}}}^{\pm}\,\circ\,T_{\theta,{\mbox{\tiny{\rm can}}}}\,\circ\,\rho_\nu(X)\,\circ\,T_{\theta,{\mbox{\tiny{\rm can}}}}^{-1}\,\circ\,\CF_{\mbox{\tiny{\rm can}}}^{\pm}
\;=\;T_{\theta,{\mbox{\tiny{\rm can}}}}\,\circ\,\CF_{+\omega}\,\circ\rho_\nu(X)\circ\,\CF_{+\omega}\,\circ\,T_{\theta,{\mbox{\tiny{\rm can}}}}^{-1}\;.
\end{equation*}
A direct computation then yields
\begin{lem}\label{LSRHO}
With $\nu=\frac{\theta}{i}$, one has:
\begin{eqnarray*} 
\CF_{+\omega}\rho_\nu(H)\CF_{+\omega}\,u(a,v,z)&=&iz\,u(a\,,\,v\,,\,z)\;;\\
\CF_{+\omega}\rho_\nu(w)\CF_{+\omega}\,u(a,v,z)&=& e^{i\partial_z}\,\left[i\cosh\left(\theta a\right)\omega(v,w)\,-
\frac{1}{{\theta}}\sinh\left(\theta a\right)\,\partial_w\,\right]u(a\,,\,v\,,\,z)\;;  \\
\CF_{+\omega}\rho_\nu(E)\CF_{+\omega}\,u(a,v,z)&=&-\frac{i}{{\theta}}\sinh\left({2\theta} a\right)\,e^{2i\partial_z}\,u(a\,,\,v\,,\,z)\;.
\end{eqnarray*}
\end{lem}
Intertwining the latter by the (partial) Fourier transform (\ref{partialFourier}), we obtain the following expressions:
\begin{eqnarray*} 
\CF_\z\CF_{+\omega}\rho_\nu(H)\CF_{+\omega}\,u(a,v,\xi)&=&-\partial_\xi\,\hat{u}(a\,,\,v\,,\,\xi)\;;\\
\CF_\z\CF_{+\omega}\rho_\nu(w)\CF_{+\omega}\,u(a,v,\xi)&=& e^{-\xi}\,\left[i\cosh\left(\theta a\right)\omega(v,w)\,-
\frac{1}{{\theta}}\sinh\left(\theta a\right)\,\partial_w\,\right]\hat{u}(a\,,\,v\,,\,\xi)\;;  \\
\CF_\z\CF_{+\omega}\rho_\nu(E)\CF_{+\omega}\,u(a,v,\xi)&=&-\frac{i}{{\theta}}\sinh\left({2\theta} a\right)\,e^{-2\xi}\,\hat{u}(a\,,\,v\,,\,\xi)\;.
\end{eqnarray*}
Also,
\begin{eqnarray*}
\CF_\z\rho_\nu(w){u}&=&e^{-a}\left[\cosh\left(\theta \xi\right)\partial_w\hat{u}+
\frac{i}{{\theta}}\omega(v,w)\,\sinh\left(\theta \xi\right)\right]\hat{u}\;;\\
\CF_\z\rho_\nu(H){u}&=&-\partial_a\hat{u}\;;\\
\CF_\z\rho_\nu(E){u}&=&-\frac{i}{{\theta}}\sinh\left({2\theta} \xi\right)
e^{-2a}\hat{u}\;.
\end{eqnarray*}
Noting the perfect symmetry in variables $a$ and $\xi$ between these expressions, we choose to
work at the level of the space $\tilde{\S}:=\{(a,u,\xi)\}$. 
\begin{prop}  \label{quadra}  Set
\begin{equation*}
\CT\;:=\;\CF_\z^{-1}\circ\CM_{\sqrt{\mbox{\rm Jac}_{\phi_\theta^{-1}}}}\circ(\phi^{-1})^\star_\theta\;.
\end{equation*}
Then, for all $\Omega\in\R$, one has:
\begin{eqnarray}\label{vulklapla}
&& \Delta^{LS} =  \CT^{-1}\,\big(\,\Omega^2\,\CF_{\mbox{\tiny{\rm can}}}^{\pm}\,\Delta^R\,\CF_{\mbox{\tiny{\rm can}}}^{\pm}\,+\,\Delta^R\,\big)\,\CT\;  \\ \nonumber
 &=& \Omega^2\partial^2_\xi+\partial^2_a +\big(\frac{\Omega^2}{\theta^2}e^{-2\xi}\sinh^2(\theta a)+e^{-2a} \cosh^2(\theta\xi)\big)\Delta_V\\ \nonumber
&-&\frac{2i}{\theta}(\Omega^2 e^{-\xi}\sinh{2\theta a} -e^{-a}\sinh{2\theta \xi} )\mathfrak{E}_V+\dim(\S)\,(\Omega^2\partial_\xi+\partial_a)\\ \nonumber
&-&\frac{1}{\theta^2}\big(\Omega^2 e^{-4\xi}\sinh^2(2\theta a) +e^{-4a}\sinh^2(2\theta \xi)\big)\\ \nonumber 
&-&\big(\Omega^2 e^{-2\xi}\cosh^2(\theta a)+e^{-2a}\frac{1}{\theta^2}\sinh^2(\theta \xi)\big)|v|^2\;;
\end{eqnarray}
where $\mathfrak{E}_V$ denotes the Euler vector field on $V$:
\begin{equation*}
\mathfrak{E}_V\;:=\;\sum_k(\omega(v,e_k)\partial_{e_k} + \omega(v,f_k)\partial_{f_k})\;;
\end{equation*}
where $\Delta_V$ denotes the Laplacian on $V$:
\begin{equation*}
\Delta_V\;:=\;\sum_k(\partial^2_{e_k} +\partial^2_{f_k})\;;
\end{equation*}
and where 
\begin{equation*}
|\,v\,|^2\;:=\;\sum_k(\omega^2(v,e_k)+\omega^2(v,f_k))\;.
\end{equation*}
\end{prop}

\section{Scalar Field Theory on SSS}

In this section we give first the general form of a scalar 
field theory on SSS space with polynomial action, then derive a Moyality 
principle for that action (at leading order),
assuming that the propagator at high energy behaves
like the flat space one, which is a reasonable assumption.

\subsection{The Action}

The quantum field theory of a scalar in a symmetric space is defined by the generic action
\bea
S=\int \phi \Delta^{LS}\phi +V ( \phi)_\star  \, .
\eea
where $V$ is some $\star$ polynomial in $\phi$ and $\Delta_{LS}$ is 
the vulcanized Laplacian defined in (\ref{vulklapla}).

The flat Fourier transform can be written as
\bea
{\cal F}_{+\omega}(f)=\delta\star^0_{\theta} f  \qquad {\cal F}_{-\omega}(f)=f\star^0_{\theta} \delta \, .
\eea
with $\delta \star^0_{\theta} \delta=1$. 

This property implies that the action is LS covariant. To see this consider the product
\begin{eqnarray}
&&{\cal F}^{-}_{\mathrm{can}} \phi \star {\cal F}^{+}_{\mathrm{can}} \phi=
T( T^{-1} {\cal F}^{-}_{\mathrm{can}} \phi \star^{0}_{\theta}   T^{-1}{\cal F}^{+}_{\mathrm{can}} \phi  ) 
=T({\cal F}_{-\omega} T^{-1}\phi \star^{0}_{\theta} {\cal F}_{+\omega} T^{-1} \phi )
\nonumber\\
&&=T(T^{-1}\phi \star^{0}_{\theta} T^{-1} \phi )= \phi \star \phi \, .
\end{eqnarray}
The same transformation for the quadratic part of the action holds, as
\bea
{\cal F}^{-}_{\mathrm{can}} \phi \star \Delta^{LS} {\cal F}^{+}_{\mathrm{can}} \phi=
{\cal F}^{-}_{\mathrm{can}} \phi \star {\cal F}^{+}_{\mathrm{can}} 
\Delta^{LS}  \phi=  \phi \star  \Delta^{LS}  \phi
\eea
and $ {\cal F}^{+}_{\mathrm{can}} $ is an involution.
Note that the metric (derived from the Laplacian) is in these coordinates
\bea
g^{\mu \nu}=
\begin{pmatrix} 1  & -2z & x^1 & x^2 \\
                         &  1+ 4z^2 +\frac{ [(x^1)^2+(x^2)^2]}{4} & \frac{ x^2}{2}-2zx^1 & -\frac{ x^1}{2} -2zx^2 \\
                         & & 1+(x^1)^2 & x^1x^2 \\
                         & & & 1+(x^2)^2
\end{pmatrix}
\eea
where $\mu,\nu=a,z, x^1,x^2$.

\subsection{Moyality principle}
\label{moyality}

In order to carry out a study of the renormalizability of this theory we need to introduce the $p$-point kernel 
corresponding to a $p$-point vertex. To compute it, start from
\bea
\mathrm{Tr}(f_p\star f_{p-1}\star \dots \star f_{1})=
\int \prod_{i=1}^p dx_i \prod_{i=1}^pf(x_i)~\mathrm{Tr}(\delta_{x_p}\star \delta_{x_{p-1}}\star \dots \star \delta_{x_1}),
\eea
where we have used $f(x)=\int dx_1 f(x_1)\delta_{x_1}(x)$ and the linearity of the $\star$ and of the trace.
The $p$ point vertex kernel is thus
\bea\label{pkernel}
  \K^p(x_p,\dots,x_1)=\mathrm{Tr}(\delta_{x_p}\star \delta_{x_{p-1}}\star \dots 
  \star \delta_{x_1}) .
\eea

As the trace is cyclic so is the $p$-point kernel. Note that if the $\star$ product is tracial the $3$-point kernel is identical with the kernel of the product between (denoted $K(x,y,z)$)
\bea
\mathrm{Tr}(\delta_{x_3}\star \delta_{x_2}\star \delta_{x_1})=
\int dx \delta_{x_3}(x)\int dydz K(x,y,z)\delta_{x_2}(y)\delta_{x_1}(z).
\eea

Let $G$ be a Feynman graph. Any reasonable scale behavior of the propagator 
will lead to locality in the ultraviolet 
region (as it behaves like a flat propagator). We will consider for the remainder of this section the limit in which the propagator is ultralocal, that is $C(x,y)=\delta(x-y)$. Indeed at short distance
this is expected to be the right approximation even in curved space to study the 
"Moyality" of the counterterms \cite{Gurau:2005gd,Rivasseau:2007ab}

In order to rewrite the vertex contribution of the graph we chose a rooted tree in the graph. We assign a total ordering of the tree lines coresponding to turning around the tree in the trigonometric way.  We introduce the first topological operation, the tree line reduction. At any step we contract the lowest  line (in the total ordering) $\ell=(i,j)$ connecting the root vertex $V$ at $i$ to another vertex $v$ at $j$.

To evaluate the integral corresponding to this tree line we cyclically turn the root vertex so that $i$ becomes the first point on $V$. We also turn $v$ so that $j$ becomes the last point on $v$. The integral associated to this line is:
\bea
&&\int dz dt~\K^{p+1}(x_p,\dots, x_1,z)\delta(z-t)\K^{q+1}(t,y_{q},\dots,y_1)=
\int dz dt du dv ~(\delta_{x_p}\star \dots \star \delta_{x_1})(u)
\nonumber\\
&&\delta_{z}(u)
\delta(z-t)\delta_{t}(v)(\delta_{y_q}\star\dots\star\delta_{y_1})(v)
=\K^{p+q}(x_p,\dots,x_1,y_q,\dots,y_1) .
\eea
where we have used the associativity and the traciality of the product.

After reducing the tree we end up with only the root vertex with some points on the root still connected by loop propagators. This vertex is called the {\it rosette} of the graph. Denote the points on the rosette by $y_p,\dots y_1$. They divide in loop half lines $(y^\ell_i, y^\ell_j)=\ell$, and true external points $y^e$. We have thus proved the lemma

\begin{lem}
  The vertex contribution at the amplitude of a graph is given by
  \bea
   A(y^e,\dots, y^e)=\int \prod_{\ell}dy^{\ell}_i dy^{\ell}_j 
    \delta(y^{\ell}_i-y^{\ell}_j)\K^p(y_p,\dots, y_1),
  \eea
   where $\K^p$ is the $p$-point kernel corresponding to the rosette.
\end{lem}

We now define the second topological operation. Take a loop line $\ell'$ on the rosette (if it exists) such that the two endpoints of the line are nearest neighbors on the rosette. We can cyclically permute the rosette to set the first two points to be the end points of the line. The contribution of such a configuration is:
\bea
&&\int dy_1 dy_2 ~\K^{p}(y_p,\dots,y_2, y_1)\delta(y_2-y_1)=
\nonumber\\
&&\int dy_1 dy_2 du~ (\delta_{y_p}\star\dots\star\delta_{y_3})(u)
(\delta_{y_2}\star\delta_{y_1})(u)\delta(y_2-y_1).
\eea

Using the product $\ref{starcan}$ we have
\begin{eqnarray}
&\int dy_1 dy_2~(\delta_{y_2}\star\delta_{y_1})(u)\delta(y_2-y_1)=
\int dy~K(u,y,y)= \nonumber \\
&\frac{1}{\theta^{\dim \S}}
\int_{-\infty}^{\infty}da_ydx_ydz_y
\sqrt{\cosh[2(a_u-a_y)]\cosh[2(a_y-a_u)]} \nonumber\\
&\times\,[\cosh(a_u-a_y)\cosh(a_y- a_u)\,]^{\frac{\dim\S-2}{2}}= \nonumber\\
&\frac{1}{\theta^{\dim \S}}\int_{-\infty}^{\infty}dx_y\int_{-\infty}^{\infty}dz_y
\int_{-\infty}^{\infty}da_y \cosh(2a_y)[\cosh(a_y)]^{\frac{\dim\S-2}{2}},
\end{eqnarray}
where we translated $a_y$ by $a_u$. The above integral is an infinite constant.

Thus the second topological operation for a line $\ell'$ such that its endpoints are neighbors on the rosette simply erases its endpoints on the rosette and multiplies the latter by this infinite constant. 

For a planar graph with only one face broken by external points we can iterate this operation for all loop lines. We have thus the lemma
\begin{lem}
The factor of a planar one broken face $p$ point graph is the $p$ point kernel times an infinite constant.
\end{lem}

This is the precise meaning of the Moyality of the theory: an arbitrary planar one broken face graph can be renormalized (if needed) by a Moyal counterterm.

\section{Extended Algebra and Metaplectic construction}
The relevance of the Langmann-Szabo self-dual propagator and its spectral analysis underlying the vulcanization process within the flat context may be independently interpreted in terms of the
classical metaplectic representation of the Heisenberg group. Indeed, we start by considering the (commutative
or not) flat configuration space $\R^d:=\{(x^1,...,x^d)\}$ as a Lagrangian subspace of 
a co-adjoint orbit $\CO\simeq T^\star(\R^d)$ of the Heisenberg group ${\bf H}_d$ in the dual $\h^\star_d$
of its Lie algebra $\h_d$. In this framework, 
a natural representation Hilbert space $(\CH= L^2(\R^d)\,,\,\rho_\hbar)$ is canonically defined (through Kirillov's theory) as well as the following quantization rule:
\begin{eqnarray}\label{WQ}
\mbox{\rm Op}:\CS(\CO)&\longrightarrow&\CB(\CH):u\mapsto\mbox{\rm Op}(u)\\
\mbox{\rm Op}(u)\varphi&:=&j^\star(\, u\star^0_\theta\pi^\star\varphi\,)\;,
\end{eqnarray}
where $\pi:\CO=T^\star(\R^d)\to\R^d$ denotes the natural projection, $j:\R^d\to T^\star(\R^d)$
the null section, where $\star^0_\theta$ denotes the Weyl product on $\CO=T^\star(\R^d)$ and 
$\CB(\CH)$ denotes the bounded operators on $\CH$. The symplectic linear group $\mbox{\rm Sp}(d,\R)$
naturally acts by automorphisms on the Heisenberg group ${\bf H}_d$, inducing an action on $\h^\star_d$ and on $\CO$ by restriction. Note that the Moyal-Weyl product is characterized as the only star-product on $\CO$ that
is invariant under the action of the linear canonical transformation group\footnote{ In particular, starting from a noncommutative configuration Moyal-space $\R^d_\theta$ rather than a commutative one does
not affect in any respect the present discussion.} $\mbox{\rm Sp}(d,\R)\times{\bf H}_d$. The metaplectic
group/representation can now be defined as follows \cite{vanH}. One first observes that the Levy factor of 
the Lie algebra of derivations of $\h_d$ that vanish on its center is isomorphic to the Lie algebra $\mathfrak{sp}(d,\R)$ of $\mbox{\rm Sp}(d,\R)$. The Stone-von Neumann theorem\footnote{It characterizes a
unitary irreducible representation of the Heisenberg group by its character of the center.} then implies that 
every element $A$ of the corresponding analytic subgroup of automorphisms of $\h_d$ acts (up to a sign) on the Hilbert space $\CH$ via the representation $\mu$ through the defining formula:
\begin{equation*}
\mu(A)\circ\rho_\hbar\circ\mu(A^{-1})\;:=\;\rho_\hbar\circ A\;.
\end{equation*}
It turns out that the vulcanized propagator $\Delta_0+\Omega^2|x^2|$ is nothing else than the operator in $\CH$ associated to
a specific element $Z_\Omega$ of $\mathfrak{sp}(d,\R)$ by the infinitesimal metaplectic representation (\cite{Folland}, p. 186):
\begin{equation*}
\Delta_0+\Omega^2|x^2|\;=\;{4\pi i}\,\mu_{\star e}(Z_\Omega)
\end{equation*}
where 
\begin{equation*}
Z_\Omega\;:=\;\left( \begin{array}{cc}0&\frac{\Omega^2}{8\pi^2}I\\ I&0 \end{array}\right)\;.
\end{equation*}
In particular, the spectral analysis of our vulcanized propagator can be performed purely in classical representation theoretical terms. For instance, Mehler's kernel can be simply derived from the expression
of (analytic continuation) of $\mu\exp(tZ_\Omega)$.

 Alternatively, in order to compute the operator semigroup $e^{t(\Delta_0+\Omega^2|x^2|)}$, one may also use the well-known explicit expressions for the Moyal star-exponentials of quadratic Hamiltonians (see e.g. \cite{OMYY}). Indeed, 
extending the above quantization rule (\ref{WQ}) to polynomial symbols the operator semigroup may be regarded as 
\begin{equation*}
e^{t(\Delta_0+\Omega^2|x^2|)}\varphi\;=\;j^\star(\, e_\theta^{t\,\lambda_{Z_\Omega}}\star^0_\theta\pi^\star\varphi\,)\;,
\end{equation*}
where $\lambda_X$ denotes the classical (quadratic) Hamiltonian function on $\CO$ associated to the element 
$X\in\mathfrak{sp}(d,\R)$ and where the star exponential is formally defined as 
\begin{equation*}
e_\theta^{\,f}\;:=\;\sum_{k=0}^\infty\frac{1}{k!}\,(f\,\star^0_\theta ...\star^0_\theta \,f)\qquad(k\;\mbox{\rm times})\;.
\end{equation*}
We therefore observe that a natural geometrical framework to implement the Langmann-Szabo duality consists in implementing our situation within a purely representation theoretical context where
the configuration space plays the role of polarized orbit. In view of passing to a general curved situation, a first essential ingredient is therefore a notion of `extended' algebra that will play to role of the Heisenberg
algebra in the flat case as discussed above. It turns out that the quantum moments $\rho_\nu(X)$ together 
with their LS-dual elements $\CF_{+\omega}\rho_\nu(X)\CF_{+\omega}$ ($X\in\s$) generate a {\sl finite dimensional} Lie algebra.

 Now, back to our curved situation, the consideration made in Section \ref{LSDUAL} naturally leads to considering the algebra  generated by the following differential operators acting  on $C^\infty(\tilde{\S})$:
\begin{eqnarray*}
X^0\;:=\;\CF\rho_\nu(X)\CF^{-1} \;\;\mbox{\rm and }\;X^1\;:=\;\CF\CF_{+\omega}\rho_\nu(X)\CF_{+\omega}\CF^{-1}\;\;\mbox{\rm where }\;X\in\s\;.
\end{eqnarray*}
It turns out that the above operators generate a finite dimensional Lie algebra, that consists in an analog 
of the Heisenberg algebra in the present curved context.
Indeed, one first sets 
\begin{eqnarray*}
w^{10}&:=&-\theta e^{-a}\left(\sinh(\theta\xi)\partial_w+\frac{i}{\theta}\omega(v,w)\cosh(\theta\xi)\right)\qquad(w\in V)\\
E^{01,1}&:=&ie^{-2\xi}\cosh(2\theta a)\\
w^{01}&:=&-\theta e^{-\xi}\left(i\sinh(\theta a)\omega(v,w)-\frac{1}{\theta}\cosh(\theta a)\partial_w\right)\qquad(w\in V)\\
E^{10,0}&:=&-ie^{-2a}\cosh(2\theta\xi)\\
E^{\epsilon,\epsilon'}&:=&\frac{i}{4\theta^2}(\theta^2+\epsilon\epsilon')e^{\epsilon(\theta-\epsilon)\xi+\epsilon'(\theta-\epsilon')a}\qquad(\epsilon,\epsilon'\in\{1,-1\}\;,\theta\neq0)\;.\\
\end{eqnarray*}
One then  computes:
\begin{eqnarray*}
\left[\,H^0\,,\,w^1\,\right]&=w^{01}, \quad \quad \left[\,H^1\,,\,w^0\,\right]&=w^{10},  \\
\left[\,H^0\,,\,w^{10}\,\right]&=w^{10}, \quad \quad  \left[\,w^{01}\,,\,v^1\,\right]&=\omega(w,v)\,E^{01,1} ,\\
\left[\,H^0\,,\,w^{01}\,\right]&=\theta^2w^1, \quad \quad     \left[\,w^{10}\,,\,v^0\,\right]&=\omega(w,v)\,E^{10,0},\\
\left[\,H^1\,,\,w^{01}\,\right]&=w^{01}, \quad \quad    \left[\,w^{10}\,,\,v^{10}\,\right]&=-\theta^2\omega(w,v)E^0,\\
\left[\,H^1\,,\,w^{10}\,\right]&=\theta^2w^0, \quad \quad \left[\,w^{01}\,,\,v^{01}\,\right]&=-\theta^2\omega(w,v)E^1,
\end{eqnarray*}
\begin{eqnarray*}
\left[\,w^{0}\,,\,v^{1}\,\right]&=&\omega(w,v)\,(E^{++}+E^{--}+E^{+-}+E^{-+})\qquad(\ast) ,
\end{eqnarray*}
\begin{eqnarray*}
\left[\,H^0\,,\,E^1\,\right]&=2E^{01,1} , \quad \quad \left[\,H^1\,,\,E^0\,\right]&=-2E^{10,0},  \\
\left[\,H^0\,,\,E^{01,1}\,\right]&=2\theta^2E^1, \quad \quad  \left[\,H^1\,,\,E^{01,1}\,\right]&=2E^{01,1} ,  \\
 \left[\,H^0\,,\,E^{10,0}\,\right]&=2E^{10,0} , \quad \quad \left[\,H^1\,,\,\,E^{10,0}\right]&=-2\theta^2E^0,
\end{eqnarray*}
\begin{eqnarray*}
\left[\,H^0\,,\,E^{++}\,\right]&=(1-\theta)E^{++},\quad  \quad  \left[\,H^1\,,\,E^{++}\,\right]&=(1-\theta)E^{++},\\
\left[\,H^0\,,\,E^{+-}\,\right]&=(1-\theta)E^{+-},\quad  \quad \left[\,H^1\,,\,E^{+-}\,\right]&=(1+\theta)E^{+-},\\
\left[\,H^0\,,\,E^{-+}\,\right]&=(1+\theta)E^{-+},\quad  \quad  \left[\,H^1\,,\,E^{-+}\,\right]&=(1-\theta)E^{-+},\\
\left[\,H^0\,,\,E^{--}\,\right]&=(1+\theta)E^{--},\quad  \quad   \left[\,H^1\,,\,E^{--}\,\right]&=(1+\theta)E^{--}.
\end{eqnarray*}
From the above brackets, one observes that Jacobi identity implies that the vector space
\begin{equation*}
\g^{\mbox{\rm ext}}\;:=\;\span\{X^0,X^1,w^{10},w^{01},E^{10,0},E^{01,1},E^{\epsilon,\epsilon'}\}_{X\in\s,w\in V}
\end{equation*}
closes as a finite dimensional Lie algebra.  Note that:
\begin{enumerate}
\item[-] the structure of the Lie algebra $\g^{\mbox{\rm ext}}$ slightly depends on the value of the real parameter $\theta$.
\item[-] The structure equation above labeled by $(\ast)$ is actually not singular for the value zero of the
parameter. Indeed, it may be re-expressed as 
\begin{equation*}
\left[\,w^{0}\,,\,v^{1}\,\right]\;=\;\omega(w,v)\,i \,e^{-a-\xi}\,\left(\cosh(\theta a)\cosh(\theta\xi)+\frac{1}{\theta^2}
\sinh(\theta a)\sinh(\theta\xi)
\right)\;.
\end{equation*}
\item[-] For generic values of $\theta$ and when $\dim\s\geq4$, the dimension is given by
\begin{equation*}
\dim\g^{\mbox{\rm ext}}\;=\;4\dim\s+2\;=\;2(2\dim\s-1)+4\;=\;2\dim\g^c+4\;;
\end{equation*}
where $\g^c$ denotes the transvection algebra of the contracted symmetric space $(\S,s)$.
\item[-] In the two-dimensional case ($\dim\s=2$), the Lie algebra structure degenerates as
\begin{eqnarray*}
\left[\,H^0\,,\,E^1\,\right]&=2E^{01,1} , \quad \quad \left[\,H^1\,,\,E^0\,\right]&=-2E^{10,0},\\
\left[\,H^0\,,\,E^{01,1}\,\right]&=2\theta^2E^1 , \quad \quad    \left[\,H^1\,,\,E^{01,1}\,\right]&=2E^{01,1},\\
\left[\,H^0\,,\,E^{10,0}\,\right]&=2E^{10,0}   , \quad \quad     \left[\,H^1\,,\,\, E^{10,0}\right]&=-2\theta^2E^0.
\end{eqnarray*}
\end{enumerate}
In the two-dimensional case the (6-dimensional) Lie algebra $\g^{\mbox{\rm ext}}$ turns out 
to be isomorphic to the `double' of the transvection algebra of the contracted symmetric space $(\S,\omega,s)$. Indeed, the parts $\{H^0, E^1,E^{01,1}\}$ and $\{H^1,E^0,E^{10,0}\}$ respectively generate
supplementary  subalgebras that are both isomorphic to $\g^c$. Moreover, $\g^c$ carries a natural symmetric space structure resembling to the classical `exchange case' in the semisimple theory \cite{Berger}:
\begin{equation*}
\sigma^c(H^0):=H^1\;,\;\sigma^c(E^0):=E^1\;,\;\sigma^c(E^{01,1}):=E^{10,0}\;,(\sigma^c)^2:=\mbox{\rm id}\;.
\end{equation*}
An analogous discussion can be performed in the general case. We postpone to a subsequent work the detailed analysis of the Kirillov unireps associated with the relevant co-adjoint orbits.

\section{Conclusion}

We have defined the analog of the Grosse-Wulkenhaar model on SSS.
The next steps should be to renormalize and explore the physics of these models.
Renormalization requires

\begin{itemize}

\item A "scale analysis" or more precisely a spectral analysis of the propagator, 
to define renormalization group steps.

The propagator $C= 1/Q$ is the inverse of the quadratic part of the action $Q$.
The Schwinger representation $C = \int_0^{\infty}  e^{- tQ} dt$ is convenient to define
a geometric series of scales such as $C = \sum C^i$, $C^i = \int_{M^{-i}}^{M^{-(i-1)}}  e^{- tQ} dt$.
In an appropriate representation one should derive explicit bounds that 
capture both the short distance behavior (governed by the Laplacian
hence expected similar to the heat kernel behavior of ordinary commutative space) 
and the long distance behavior that depends on the vulcanization terms.
This representation could be either direct space, coherent states (also called "matrix representation")
or some mixed representation like the partial Fourier transform used in (\ref{partialFourier}) and Proposition \ref{quadra}.

\item These bounds should be combined with the vertex kernel (computed in the same representation)
in order to establish power counting for Feynman graphs. This power counting is dimension- and model-dependent.
In dimension four the theory should be renormalizable. This means that the
only amplitudes which diverge when the difference between their internal and external scales increases
can be written as a local part of the form of the initial action
plus a remainder which no longer diverges. For the coupling 
constant and mass renormalization this step should be 
a direct consequence of the scale analysis and of the Moyality
principle of section \ref{moyality}. However the so-called wave function renormalization
is a bit more subtle. It requires an extension
of that Moyality principle to "second order"
for the two point function. Indeed beyond the zeroth order which corresponds to mass renormalization
one needs to prove that the second order approximation recovers exactly the coefficients
of the vulcanized propagator, namely renormalizes the coefficient of the Laplacian (this is usually
absorbed in a rescaling of the field variable), and the coefficient of the 
vulcanization term.

\item Renormalization flows should be computed to check whether absence
of Landau ghosts survive on these non flat SSS backgrounds. If this is the case
(as we a priori expect) the non perturbative construction of the theory 
should follow. 

\end{itemize}

This renormalization program and the mathematical relation of LS duality to metaplectic representations
will be explored in future publications.

\end{document}